# Realising the full potential of X-ray Astronomy in the UK
*To be submitted to A&G (Aug 2017)*

**Abstract**
_______________________________

X-ray astronomy is our gateway to the hot universe. More than half of the baryons in the cosmos are too hot to be visible at shorter wavelengths. Studying the extreme environments of black hole and neutron star vicinities also requires X-ray observations. With the successful launch of ISRO's *AstroSat* in 2015, and a few transformative results that emerged from JAXA's *Hitomi* mission in 2016, a new window has been opened into high sensitivity fast timing and high X-ray spectral resolution. Together with upcoming all-sky survey missions expected very soon, X-ray astronomy is entering a new era of parameter space exploration. The UK has been at the forefront of this field since the 1970s. But flat cash science budgets, compounded with the rising costs of cutting-edge space missions, imply inevitably diminishing roles for the UK in terms of both payload development and science exploitation in the future. To review the novel science possibilities enabled by recent and upcoming missions, and to discuss how to pave the way forward for X-ray astronomy in the UK, a specialist RAS discussion meeting was held in London on Feb 10 2017, summarised herein. A consolidated effort by the community to come together and work cohesively is a suggested natural first step in the current climate.
_______________________________

X-ray astronomy is a little over 50 years young, having begun in earnest in 1962 with the discovery of extra-solar X-ray sources (Giacconi et al. 1962). As compared to a more mature field such as modern optical astronomy, the development of which may be traced back to the first refracting telescopes in the early 1600s, X-ray astronomy is still only in its tweens. And it is experiencing growing pains.

There is an explosion of data and results over the entire X-ray spectral regime, extending at least three decades in energy from ~0.1-100 keV. Driving this growth are a number of international observatories probing deeper and wider in imaging, spectroscopy and, soon, polarimetry. February of 2016 saw the launch of the JAXA-led *Hitomi* space mission, opening up an era of high spectral resolution X-ray astronomy with microcalorimeters. India's *AstroSat* mission had been launched less than 5 months prior to that, rejuvenating the possibilities for high sensitivity rapid X-ray timing. Time domain studies have been core to X-ray astronomy since the early days, and this field will undoubtedly continue to mature with the high quality of data sampling now available.

The UK has traditionally punched above its weight in terms of science return relative to the size of the X-ray community here. With the aim of providing a forum for UK astronomers to discuss the key science themes from these powerful new missions, and to discuss areas of expertise development required to best exploit the large new regions of parameter space available, a proposal was submitted soon after *Hitomi*'s launch to hold a specialist discussion meeting on the theme of 'Timing and Spectroscopy in the new era of X-ray astronomy' at the Royal Astronomical Society (RAS).

But the opacity of the atmosphere means that X-ray astronomy can only be done from space, and space can be unforgiving. Contact was lost with *Hitomi* only 37 days later, due to a chain of events culminating in an uncontrolled spin and ultimate satellite breakup (Hitomi

Experience Report). This was a particularly difficult blow for the community, given the launch failure of *Astro-E* and the coolant leak in *Suzaku*, both of which were also carrying microcalorimeters. However, a few precious observations of a handful of targets before failure ensured *Hitomi*'s legacy and showcased the spectacular potential of high resolution X-ray astronomy.

Given these developments, it was decided to hone the focus of the RAS meeting to more strategic issues. What would be the state of X-ray astronomy post *Hitomi*? Given severe funding restrictions, how can the UK continue to play a frontline role in current and future missions? The meeting was held on Feb 10 of 2017, and a summary follows below.

**The meeting**

The day began with a tribute to Neil Gehrels led by **Julian Osborne (Univ. Leicester)**. Neil passed away on Feb 6, just four days prior to the meeting. Julian commented on Neil's enormous contribution to high-energy astrophysics, and reminded us that Neil was an energetic man with a keenness to say yes to new ideas. His leadership style was inclusive and decisive, and this played a significant part in the substantial impact of the *Swift* mission. Julian's tribute was followed by a few moments of silence by the audience.

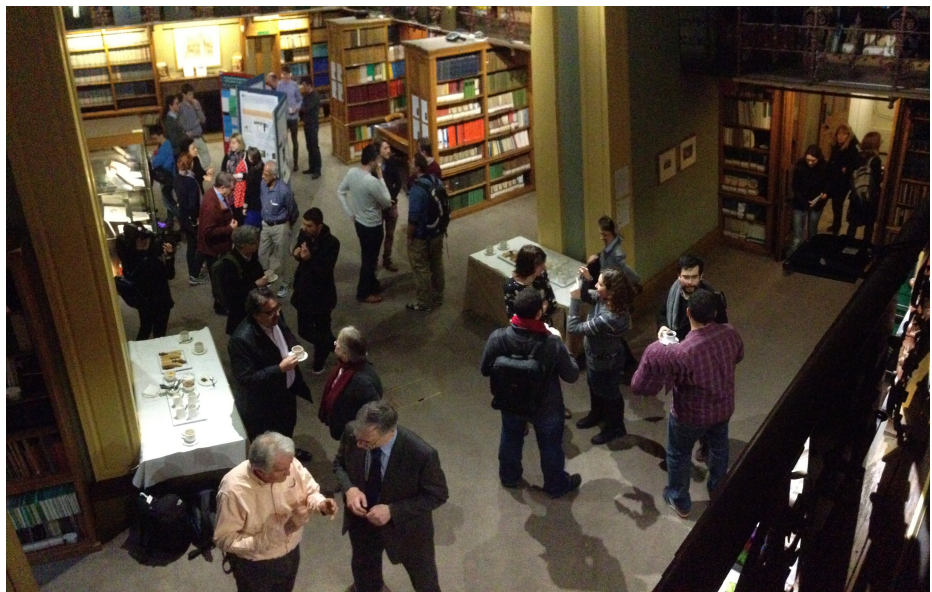

Figure 1: Participants arriving and mingling in the coffee/poster area at Burlington House on the morning of Feb 10, 2017.

A total of 10 posters were presented at the meeting, displayed throughout the day for viewing (Fig. 1). These included science results as well as presentations of upcoming missions. A list of these posters is presented in Table 1. The morning saw talks covering new prospects on both timing as well as spectroscopy, while the afternoon focused on upcoming missions culminating in a community discussion. These are described below.

| *Table 1 : Poster Presentations* | |
|---|---|
| The SVOM rapid-response multi-wavelength GRB observatory | **Julian Osborne** (Univ. Leicester) |
| XIPE: The X-ray Imaging Polarimetry Explorer | **Silvia Zane** (UCL), XIPE consortium |
| XCS, The XMM Cluster Survey – It's not just about clusters | **Julian Mayers** (Univ. Sussex) |
| The X-ray Baldwin Effect in Compton-thick AGN | **Peter Boorman** (Univ. Southampton) |
| Thermonuclear X-ray bursts in rapid succession in 4U 1636-536 with ASTROSAT-LAXPC | **Aru Beri** (Univ. Southampton) |
| X-ray and ultraviolet variability of active galactic nuclei with Swift | **Douglas Buisson** (Univ. Cambridge) |
| SMILE (Solar wind Magnetosphere Ionosphere Link Explorer): X-ray imaging of the Sun-Earth Connection | **Graziella Branduardi-Raymont** (MSSL) |
| Testing the Nature of Black Holes with X-ray Spectra | **Jiachen Jiang** (Univ. Cambridge) |
| Investigating the Evolution of the Dual AGN System ESO 509-IG066 | **Peter Kosec** (Univ. Cambridge) |
| Prospects for Galactic X-ray binaries with LSST | **Michael Johnson** (Univ. Southampton) |

**Poster abstracts and PDFs are available online at https://sites.google.com/site/ukxrayastro/**

**New prospects with X-ray timing**

Strong flux variability over a broad range of timescales is a fundamental characteristic of compact accreting sources. Hence fast timing studies are widely used in X-rays to study the nature of accretion. For 16 years starting late 1995, NASA's *Rossi X-ray Timing Explorer* (*RXTE*) mission served as the X-ray timing workhorse, enabling a huge range of discoveries with its fast response to transient events and its wide sky coverage. It was responsible for discovering high frequency oscillations in compact accretors, as well as the evolutionary connection between accreting millisecond X-ray pulsars and isolated radio pulsars. Its frequent monitoring and flexible scheduling transformed our understanding of accretion cycles in transient binaries. *RXTE*'s demise in early 2012 left the X-ray community with a significant handicap in the field of high sensitivity fast timing.

*AstroSat* (Singh et al. 2014) is, in several respects, a natural successor to *RXTE*. Launched in 2015 by the Indian Space Research Organisation (ISRO), the mission has a broadband response over at least ~0.5-100 keV. It additionally carries an ultraviolet imaging telescope (UVIT) making it a truly multiwavelength mission, and India's first astronomy-dedicated satellite. **A.R. Rao (Tata Instt. of Fundamental Research)** flew over from Mumbai to deliver a presentation on the status of *AstroSat*. All instruments are functioning nominally, although calibration and observing efficiency optimisation is ongoing. Every instrument, including UVIT, is capable of operating in event mode, though UVIT has a strict count limit and low observing efficiency. Background calibration of its wide area Sky Survey Monitor is also ongoing. Some recent results of note include serendipitously catching a black hole binary outburst on the first day of a reported outburst (Yadav et al. 2016), detection of a multitude of gamma ray bursts (Rao et al. 2016), and polarisation studies of a number of objects (e.g. Rao et al. 2016, Chattopadhyay et al. 2017, Fig. 2). The first international (10%) open call for observing was issued after the meeting in summer 2017, and we can expect a steady stream of results over the coming months and years.

ESA's premier X-ray mission in-orbit, *XMM-Newton*, continues to perform well despite having now spent almost two decades in space. **Adam Ingram (API, Netherlands)** showed the power of its fast timing mode, presenting the best evidence that quasi-periodic oscillations (QPOs) are driven by Lense-Thirring precession. This evidence comes from observations of a modulation of the fluorescence Fe Kα emission line on the QPO frequency in an outbursting Galactic black hole binary (Ingram et al. 2016). This is a new probe of the strong gravity regime in the vicinity of the black hole, which we can expect to be exploited in the future using high throughput, pile-up free instruments such as *NuSTAR*, *AstroSat*, *NICER* and *eXTP*.

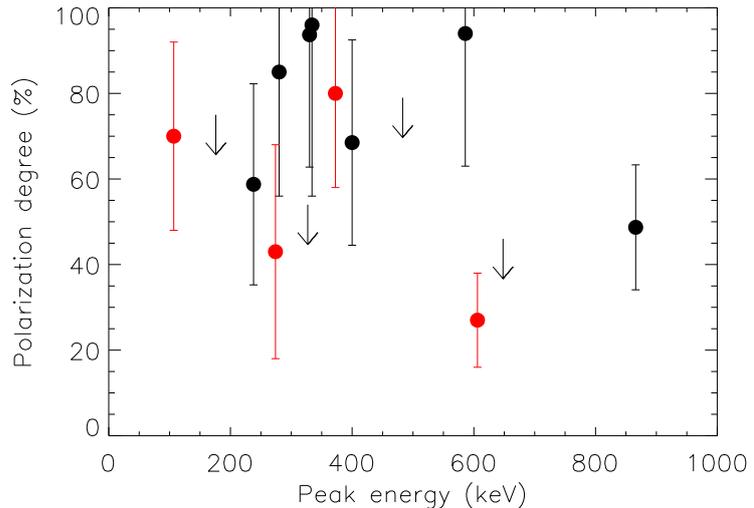

Figure 2: *AstroSat* polarisation degree measurements of Gamma Ray Bursts as a function of peak energy. The black points represent the bursts detected by the CZTI instrument and the red points denote earlier measurements by GAP/*IKAROS* and *INTEGRAL* (Chattopadhyay et al. 2017).

Accretion is an inherently broadband phenomenon, and in restricting oneself to X-rays, much information is lost. **Vik Dhillon (Univ. Sheffield)** reviewed the ongoing revolution in optical fast timing. In particular, the ULTRACAM instrument (Dhillon et al. 2007) has been at the forefront of fast optical observations coordinated with X-rays. Whereas optical delays of order ~10 seconds due to reprocessing of X-rays in the accretion disc and companion star in X-ray binaries have been known for some time (e.g. Muñoz-Darias et al. 2007), an increasing number of studies are revealing the presence of much shorter sub-second optical delays in outbursting binaries, interpreted as evidence for jet and hot flow emission (e.g. Kanbach et al. 2001, Gandhi et al. 2010, Durant et al. 2011). Such multiwavelength fast timing has been difficult to coordinate, but should become easier in the future with an increasing availability of instruments. In particular, Vik reviewed the powerful new instrument HiPERCAM (Dhillon et al. 2016), which is a simultaneous 5-channel imager capable of frame rates up to 1600 Hz. Optical fast timing is one field where the UK currently has world-leading instrument access.

At the high end of the compact object mass spectrum, active galactic nuclei (AGN) may be scaled up cousins of stellar-mass X-ray binaries. If accretion is scale-invariant, both ends of the mass spectrum can yield insight on the underlying physics, though this remains controversial. **Ian McHardy (Univ. Southampton)** presented details of large ongoing programmes to study correlated optical and X-ray variability in AGN. Intriguingly, the observations have revealed optical and ultraviolet lags (with respect to X-rays) that are significantly longer than those predicted by standard accretion disc theory. The lags scale with wavelength as expected, but are larger than model expectations by a factor of ~3 (McHardy et al. 2014, Edelson et al. 2016). Modifications to mass, accretion rate and/or disc temperature may explain this unexpected result. Coordinated long term monitoring and multiwavelength missions such as *AstroSat* can drive this field forward with the requisite sampling over months to years.

**New prospects in X-ray spectroscopy**

X-ray spectrometers now routinely sample a broad range covering one or two decades in energy, from ~0.5-100 keV and more. This is much broader than possible in the optical. However, X-ray spectrometers lag far behind the optical in terms of spectral resolution. The best instruments in this regard currently flying are the grating spectrometers on *Chandra* and *XMM-Newton* with energy resolution ($E/\Delta E$) of ~1000 around 1 keV, but only ~160 around the important Fe K$\alpha$ fluorescence emission at 6.4 keV (covered by *Chandra* alone). The X-ray community has long awaited the age of microcalorimeters in orbit, capable of far superior energy resolution at high energies. Following the launch failure of *Astro-E* and the coolant leak in the X-ray Spectrometer on board *Suzaku*, *Hitomi* was supposed to deliver on this promise.

In fact, its microcalorimeter (the Soft X-ray Spectrometer, or SXS) performed spectacularly in orbit for the first month of *Hitomi*'s life. Unfortunately, the satellite subsequently suffered a series of anomalies leading to catastrophic breakup (Hitomi Experience Report). The few observations made by SXS were carried out through a closed gate valve that reduced its sensitivity greatly. Despite this, first-light observations of the Perseus cluster probed a previously unsampled regime, achieving a spectral resolution of $\Delta E \sim 4.5$ eV at around ~6 keV. A surprisingly quiescent intracluster medium was found based upon the narrow widths of emission lines from ionised metal species, implying the presence of some mechanism to suppress turbulent motions in the cluster gas (Fig. 3; Hitomi Collaboration 2016). Another much awaited result from the mission was the search for a signature of dark matter annihilation. Whereas several recent studies have detected tentative hints of a signal around 3.5 keV (e.g. Bulbul et al. 2014, Boyarsky et al. 2014), *Hitomi*'s first-light observation ruled out the presence of a corresponding narrow emission feature at 99.7% confidence (Hitomi Collaboration 2017). A nearby charge exchange transition could potentially account for the observed excess in low spectral resolution data. **Andy Fabian (Univ. Cambridge)** reviewed *Hitomi*'s findings, telling the audience what a privilege it was to be able to work on the *Hitomi* data, and reminding everyone that it is exceedingly rare for first-light observations from any mission to result in high impact publications. This underscores the enormous untapped potential of high spectral resolution X-ray observations.

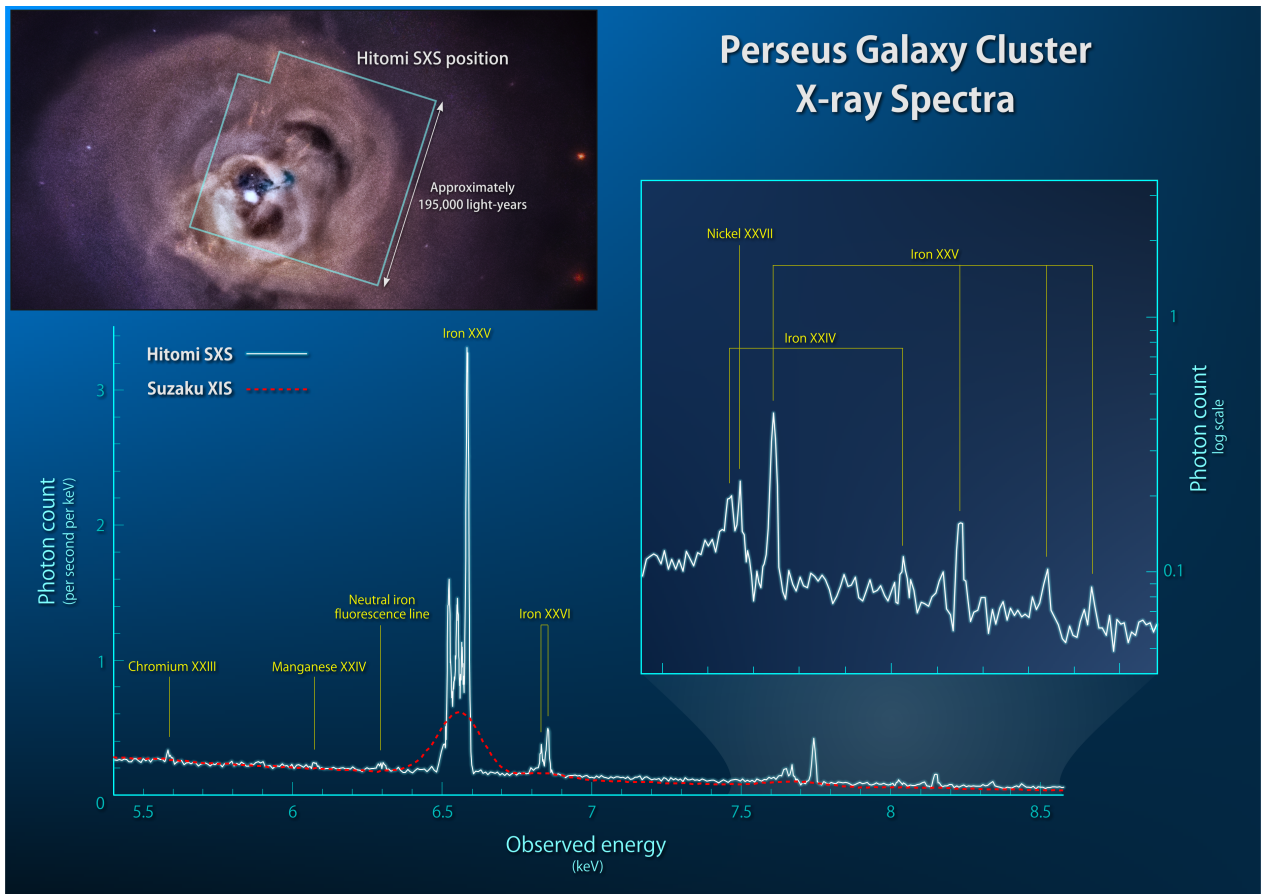

**Figure 3:** *Hitomi*'s spectrum of the Perseus cluster core. A complex of ionised species, in particular the strong Fe lines, are resolved and found to have a line-of-sight velocity dispersion of 164±10 km s$^{-1}$. This is the tightest constraint so far, far superior to CCD spectra (red dashed curve), and places strong limits on the influence of turbulence within the intracluster medium (Hitomi Collaboration 2016). Background image: NASA *Chandra* X-ray Observatory. Figure credit: NASA Goddard Space Flight Centre.

Ultraluminous X-ray sources (ULX) were long considered to be intermediate mass black holes, a missing link between X-ray binaries and AGN. This view has been largely overturned during the past few years based upon a number of observations. In particular, the discovery of pulsations in a ULX in the starburst galaxy M82 showed that at least some of these objects must be neutron stars, and hence have masses under ∼2 M$_\odot$ (Bachetti et al. 2014). Consequently, their high luminosities $L_{\text{X-ray}} \sim 10^{40}$ erg s$^{-1}$ imply that they must be accreting at super-Eddington rates. Strong outflows are expected in such a regime, and **Ciro Pinto (Univ. Cambridge)** presented recent evidence of the presence of winds in several ULXs. This work exploits the high spectral resolution of the *XMM-Newton* gratings at soft X-ray energies, finding blueshifted absorption features characteristic of outflowing gas at speeds of about 0.2$c$. Ciro further presented an orientation-dependent unification model for hyper-accreting stellar-mass compact objects, with hard X-rays emerging preferentially along the axis of an optically-thick disc wind, and softer energies dominated at increasing inclination angles due to reprocessing (Pinto et al. 2017). Such observations give important insight into the super-Eddington accretion regime.

What is the future for high spectral resolution X-ray astrophysics post-*Hitomi*? **Richard Kelley (NASA Goddard Space Flight Centre)** showed us the planned roadmap for *Hitomi*'s successor. Currently named *XARM* (*X-ray Astronomy Recovery Mission*), Richard delivered the news that the mission has approval from both JAXA and NASA, and is to be launched by ~2021[1]. It is scaled down from *Hitomi*, in that there will be no hard X-ray instrumentation. Below 10 keV, the soft X-ray telescope, imager, and microcalorimeter will be retained, making it superior to all other missions in terms of spectral resolution at energies of several keV (Fig. 4). Lessons learnt from *Hitomi* in terms of project organisation, operations, and in-orbit performance will be incorporated from the start. Richard also stressed the fact that the UK did, historically, play a leading role in building its own X-ray mission with *Ariel V* in the 1970s. That mission went on to make some fundamental discoveries including the detection of ionised Fe in intracluster plasma (Mitchell et al. 1976), and we are only just (~40 years on) beginning to probe the dynamics of this hot gas with microcalorimeters. He ended by showing that we are entering a golden era of X-ray microcalorimeters, with steady improvements in energy resolution over the past ~30 years following a progression similar to Moore's Law.

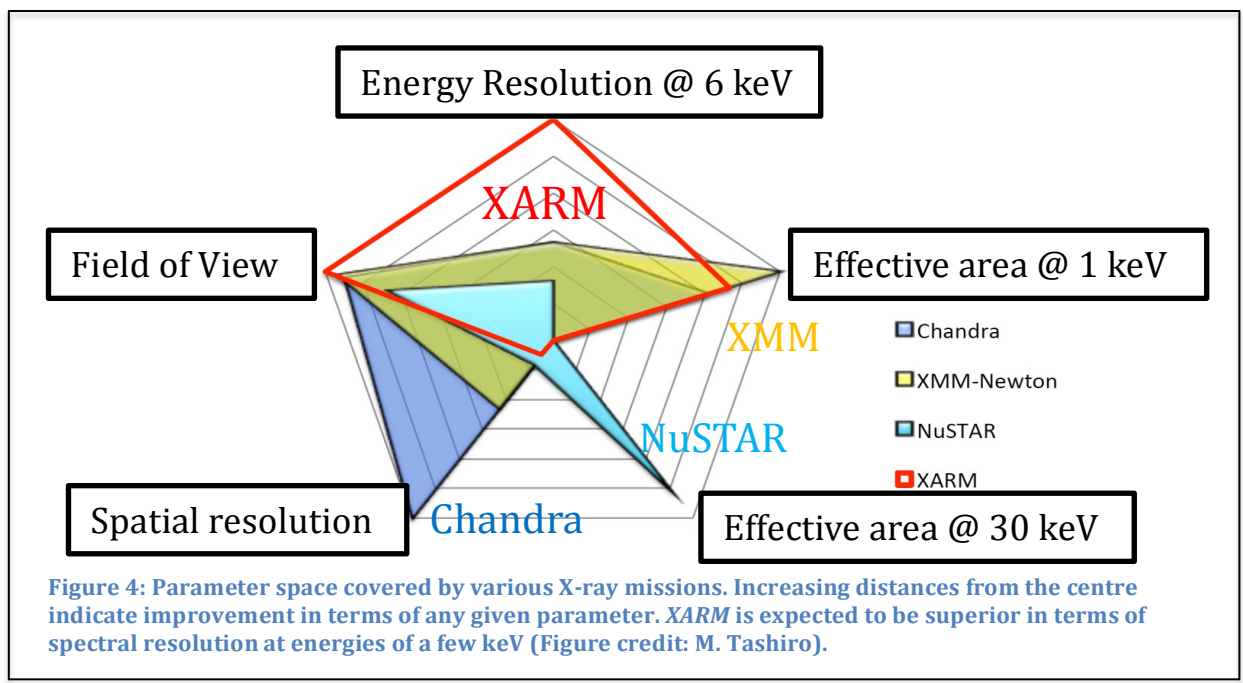

**Figure 4:** Parameter space covered by various X-ray missions. Increasing distances from the centre indicate improvement in terms of any given parameter. *XARM* is expected to be superior in terms of spectral resolution at energies of a few keV (Figure credit: M. Tashiro).

On the long-term horizon, *Athena* will be the premier X-ray observatory in terms of sensivity and high spectral resolution. **Paul Nandra (MPE)** reviewed the current status of the mission, which is expected to launch around 2028. With collecting area of 2 m² (1 keV) sampling the field of view with 5 arcsec pixels, an integral field unit of transition edge sensors with spectral resolution a factor of ~2 better than *Hitomi* and a large field of view spanning 5 arcmin (equivalent diameter), additionally combined with a wide field imager (40 armin), *Athena*'s science is expected to be transformational (Nandra et al. 2013, Willingale et al. 2013, Barret et al. 2016, Meidinger et al. 2016). The project is currently in Phase A, with mission adoption expected in 2020. The UK's main role in the mission is in the Wide Field Imager, but Paul stressed that it was surprising to see the UK not playing a stronger role in the mission as compared to other ESA partners.

**Future missions and strategies**

The afternoon session continued the theme of future synergies and strategies with presentations on wide field X-ray imaging, polarimetry and timing. **Arne Rau (MPE)** reviewed the current status of e-ROSITA on board *SRG* (*Spectrum-Roentgen-Gamma*). The instrument is ready and has been delivered for launch from Baikonur in 2018. e-ROSITA (Merloni et al. 2012) will carry out the deepest all-sky X-ray surveys so far in the energy range of 0.3-10 keV. Cosmological parameter estimation through detection and study of all massive galaxy clusters is a key science driver for the mission, but e-ROSITA will also detect more than 3 million AGN, many Galactic compact objects and ~500,000 stars. In addition, it is expected to be sensitive to a huge range of variables and transients including, for instance, about 1000 tidal disruption events over 4 years.

---

[1] In June 2017, ESA also announced its participation in the mission.

**Silvia Zane (UCL)** discussed the prospects and importance of various (concept) missions with X-ray polarimetry (*XIPE*, *eXTP*) and high resolution X-ray timing (*eXTP*, *STROBE-X*) capabilities to study neutron stars. X-ray polarimeters such as on board of the *X-ray Imaging Polarimetry Explorer* (*XIPE*), a concept mission proposed for ESA M4, and the *enhanced X-ray Timing and Polarimetry* (*eXTP*) mission, endorsed for launch in 2024-2025 by the Chinese National Academy of Sciences (CNAS), will open a completely new window to study the geometry of X-ray emitting plasmas. This is expected to lead to significant breakthroughs in our understanding of, for example, the strength and geometry of neutron star magnetospheres, particle acceleration processes, accretion disks, and the equation of state of ultra-dense matter. Silvia also highlighted the impact of a high-throughput X-ray timing mission like *eXTP* or *STROBE-X*, a NASA probe-class mission proposed within the context of the Decadal Survey 2020, to study general relativistic effects around black holes and the dense matter equation of neutron stars.

**Giorgio Matt (Roma)** followed this and discussed the design, status and key science questions to be addressed by the *Imaging X-ray Polarimetry Explorer* (*IXPE*; Weisskopf et al. 2016), not to be confused with *XIPE* above. This mission concept was selected by NASA in the Small Explorer Program (SMEX) for launch in 2020 and is currently in exploratory phase. Giorgio also stressed the immense potential of obtaining a full, comprehensive view of X-ray sources with a mission like *IXPE* that provides simultaneous energy, imaging, timing, and polarization information.

As chair of Astronomy Advisory Panel for the UK's Science and Technology Facilities Council (STFC/AAP), **Paul O'Brien (Univ. Leicester)** was invited to give us his viewpoint of how high-energy astrophysics fitted within STFC priorities and how the X-ray community could best interact with AAP. Paul reminded us that the UK share of global space science and astronomy papers far outstrips the proportion of papers in other broad comparator fields including Physics, Chemistry, Engineering and Mathematics. In other words, the UK is delivering world-class science in astronomy and space science.

The latest (2016) AAP community consultation exercise revealed that exploitation funding remains the highest concern, and this issue was raised repeatedly during the day. ESO remains the top ranked priority for the community, followed closely by high performance computing. While diversity in observing facilities is considered important, we are at the limit of a viable, world-leading astronomy programme in terms of funding. The growing scale of astronomy experiments requires astronomy grant panel (AGP) funding to increase, but the reality is that observing time available to UK astronomers continues to decline as we withdraw from observatories and as telescopes become more specialised. Support is strong for the ESA chosen missions, such as *PLATO* and *Athena*, and UKSA is likely to support whichever mission tops the M4 selection in late 2017, since the UK is involved in all the short-listed candidates. But the community is concerned that the overall level of payload funding is not in accord with our pro-rata ESA subscription spend. Paul emphasised the need for a balanced portfolio of ground-based observing facilities in order to best exploit the UK investment in existing and future space missions.

**Mike Cruise (Univ. Birmingham)** began his talk by reminding the audience that he was not presenting the official view of the UK Space Agency (UKSA). UKSA was unable to send a representative to attend the meeting. Mike gave an overview of what UKSA funding covers (in addition to our ESA subscription) with regard to new proposals for space mission involvement, including initial studies, hardware development beyond a certain technology readiness level, and in-orbit operations. UKSA is now part of BEIS (the Business, Energy and Industrial Strategy ministry). Space activity is worth £13.7 billion to the UK economy, of which about 2.2% is spent on science programmes, and about 0.1% constitutes science payload funding for ESA programmes. All proposals for new missions are discussed at the UKSA Science Programme Advisory Committee (SPAC), and the grant award is set by the Space Projects Peer Review Panel (SPPRP). Science quality is only one of the criteria assessed for any new proposal; the others being Timeliness, Economic Impact, Risk, Science Return to the UK, and Outreach.

The UK contributes about 8.7% to the ESA budget and has supported all the missions selected by ESA and supported by STFC. This "dual key" system agreed between UKSA and STFC has proven to work well. The UK has, historically, generated about twice the science outcomes that its percentage ESA budget share would suggest. However, Mike personally considers that we are currently reaping the benefits of past investments, and the UK grants for payload instruments agreed in the past few years will not buy us leading positions in future space missions. Mike ended his presentation by exhorting the audience to get on ESA committees and to help develop strong government/societal cultural support for pure science. There is no formal bar to larger space science expenditure, but it is incumbent upon the community to demonstrate its relevance.

**Community discussion**

It is clear that we face tough choices ahead. Funding is not getting any easier, and current grant levels cannot guarantee the UK a leading position in future space missions. As such, the UK X-ray community risks missing out on the ongoing global revolution in the field.

As part of the run-up to the meeting, the organisers had issued an open online survey with the following questions: "Are you satisfied with the development and support of X-ray astronomy in the UK? What more should we be doing as a community?" A couple of thought-provoking answers were received[2]. One person emphasised the need for an established presence on funding panels (e.g. fellowship and grant panels) as well as those able to lobby for government spending. This was highlighted as being important for ensuring that there are enthusiastic and experienced scientists who continue to drive innovation with the next generation of missions.

Another comment recognised the importance of *Athena* for the community, but also pointed out that the mission is still a very long way off. In addition, the comment stated, most existing X-ray facilities are well past their design life, so new opportunities must be sought, where possible. China potentially offers fruitful opportunities in the near-term horizon, including the recently announced *Einstein Probe* capable of 3600-deg² field-of-view monitoring for transients using lobster eye mirrors, set to launch in 2021. However, the fact that UKSA does not support bilateral missions with other nations excludes the UK from such dynamic space programmes. Arguably, the UK has a world lead in lobster eye technology, but we are being left behind due to the lack of collaboration possibilities, the comment concluded.

---

[2] **https://sites.google.com/site/ukxrayastro/home/community-poll/community-comments**

UK leadership in the field was brought up repeatedly during the meeting, from the UK-led *Ariel V* mission in the 1970s to participation in several ongoing major observatories. Bibliometrics (see Fig. 5) show that the UK continues to publish in a leading position in the field of X-ray astrophysics on par with other countries that have led their own missions in recent decades. This has been possible in large part due to close collaborations with other nations (either through mission team memberships at the level of individual scientists, or through ESA participation), which we must continue to nurture.

While most large missions do allow international participation through open proposal calls after an initial period of restrictive access, it must be recognised that some of the most exciting and important discoveries occur during this initial period when a new region of parameter space is being pried open. By awaiting for open calls, we not only risk missing out on participating in some of the most stimulating discoveries, there is also the danger of early mission failure. Both these aspects were exemplified by the case of *Hitomi*. Despite ESA participation, only a handful of UK-affiliated astronomers were ultimately able to participate in the high profile science papers that resulted from the early observations.

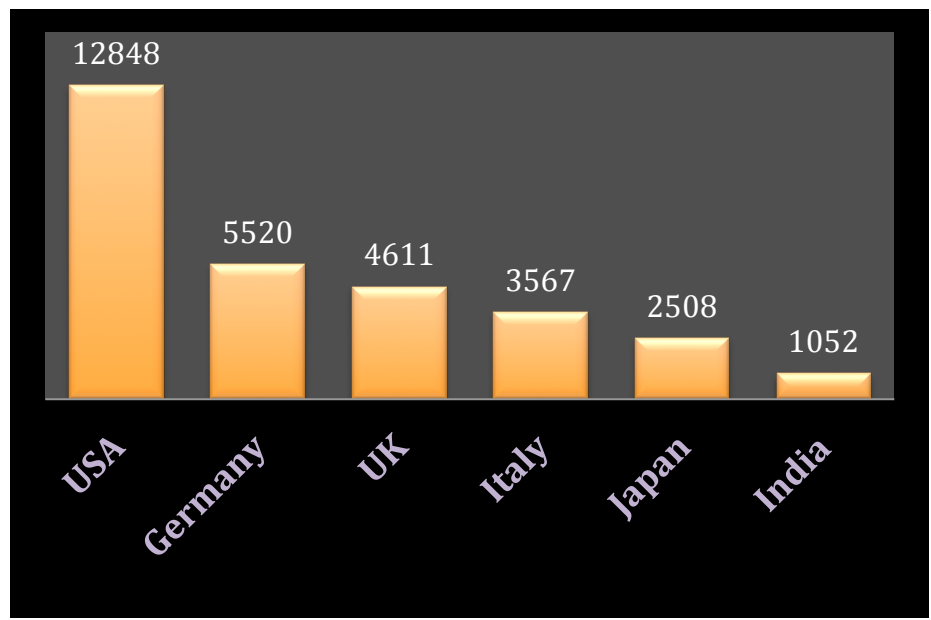

Figure 5: Statistics from NASA's Astrophysics Database Service (ADS), showing the number of astronomy publications mentioning "X-rays" over the 5-year period of 2012-2016, distributed as a function of author affiliation from six countries; retrieved on 2017 Feb 8. The UK is the only country in this list not to have launched its own X-ray mission over the past ~40 years.

With this background, the day's discussion ended with the question of "What more should we be doing as a community?" The organisers proposed that as a first step, UK X-ray astronomers should come together as a cohesive community. Only a demonstrably cohesive community can make its voice heard when it comes to major support and funding requests. This could also ensure good representation on relevant grant committees and science-policy panels.

While there was consensus that such a cohesive approach was worthwhile, there was no immediate agreement on the best way to achieve this. Creating an equivalent of the AAS/HEAD (High Energy Astrophysics Division of the American Astronomical Society) within the RAS was one suggested way to bring together UK X-ray astronomers. Although HEAD works as a coherent subgroup within the AAS, this was considered to be divisive in the UK context by several members of the audience. However, such groups are not without precedent. One example is the UK Planetary Forum (UKPF), which is affiliated to the RAS and receives RAS support for meetings and website servers. UKPF demonstrates that it is possible to strengthen and serve the needs of a specific community as a subgroup within the RAS, without being divisive. It could also help to focus our efforts to create the kinds of cultural shifts that Mike Cruise encouraged in his presentation.

Another issue raised was the need to sustainably foster the next generation of X-ray astronomers and instrument builders. However, there was a palpable mood of pessimism amongst the audience on this point, given diminishing prospects for postdoc positions (e.g. the removal of STFC postdoctoral fellowships, flat cash grant level income of STFC, uncertainty with regard to continued EU funding). Ensuring X-ray astronomy representation on funding panels was again highlighted as being important in this regard.

Finally, the possibility of bilateral buy-ins from the X-ray astrophysics community as a whole was raised. There are several near-term missions including *XARM*, *SVOM*, *Einstein Probe* and more that could be candidates for this. Another option proposed in Paul Nandra's presentation was strengthened UK support for *Athena*. All of these options were also considered to likely be prohibitively expensive given funding realities. As a counterpoint, however, the example of LSST was raised. The UK was able to recently buy in as an international partner to the LSST Consortium for ~£17 million (including funds to develop a data centre and for science exploitation), at a time of sustained flat cash funding. This was possible because of the wide remit of LSST, which encompasses an enormous range of science themes and hence garnered broad community support. If we are to make the same case for X-ray astronomy support, then we similarly need to demonstrate the relevance of X-ray studies for a variety of broad research areas within astrophysics. Examples of such synergy can be found amongst the astronomy, space plasmas and planetary science communities in ongoing joint observations of Jupiter with coordinated in-situ (*Juno*) and remote (*Chandra*/*XMM-Newton*) observations, as well as in the planning of future missions (*SMILE* was highlighted in this regard).

## Conclusions

X-ray astronomy is undergoing a global renaissance of sorts with (1) new sensitive fast-timing missions in-orbit, (2) wide-field imaging and polarimetry missions on the horizon, and (3) a recent taste of high spectral resolution studies. If the immense science output from the first-light observations of *Hitomi* is anything to go by, major breakthroughs are ahead as we open up wide new swathes of observational parameter space. If we wish to see the UK play a leading role in this ongoing renaissance, it is our responsibility to make our voices heard, and to pave the way forward despite a climate of dwindling funding possibilities. We suggest that the first step ought to be for UK X-ray astronomers to come together and to work cohesively to ensure that the importance and excitement of the science are effectively communicated not only to the relevant panels but also to the wider scientific community and to the public at large. How we achieve this is up to us.

## Authors

**Poshak Gandhi** (Univ. Southampton; p.gandhi@soton.ac.uk), **Nathalie Degenaar** (Anton Pannekoek Institute, Netherlands), **Chris Done** (Univ. Durham), **Mike G. Watson** (Univ. Leicester).

## Acknowledgements

The organisers wish to thank the RAS for the opportunity to discuss these important issues. They acknowledge research support from STFC. PG thanks Phil Charles for discussions during the initial stages of planning this meeting and for comments on an initial draft.

**UK X-ray astronomy twitter feed:** #ukxrayastro